 \documentclass[]{jfm_arxiv}

\usepackage{graphicx}
\usepackage{newtxtext}
\usepackage{newtxmath}
\usepackage{natbib}
\usepackage{hyperref}
\hypersetup{
    colorlinks = true,
    urlcolor   = blue,
    citecolor  = black,
}
\usepackage[usenames, dvipsnames]{xcolor}

\DeclareFontEncoding{LS1}{}{}
\DeclareFontSubstitution{LS1}{stix}{m}{n}
\DeclareFontEncoding{LS2}{}{\noaccents@}
\DeclareFontSubstitution{LS2}{stix}{m}{n}
\DeclareSymbolFont{SYMB}{LS1}{stixscr}{m}{n}
\DeclareSymbolFont{SYMB4}{LS1}{stixbb}{m}{it}
\DeclareSymbolFont{ARR3}{LS2}{stixtt}{m}{n}
\DeclareMathSymbol{\pentagonblack}{\mathord}{SYMB4}{"DA}
\DeclareMathSymbol{\bigblacktriangleup}{\mathord}{SYMB}{"C9}
\DeclareMathSymbol{\bigblacktriangledown}{\mathord}{SYMB}{"D3}
\DeclareMathSymbol{\mdlgblkdiamond}{\mathord}{SYMB}{"DD}
\DeclareMathSymbol{\mdlgwhtdiamond}{\mathord}{SYMB}{"DE}
\DeclareMathSymbol{\circletophalfblack}{\mathord}{SYMB}{"EA}
\DeclareMathSymbol{\circlerighthalfblack}{\mathord}{SYMB}{"E8}
\DeclareMathSymbol{\circlebottomhalfblack}{\mathord}{SYMB}{"E9}
\DeclareMathSymbol{\squaretopblack}{\mathord}{SYMB4}{"CD}
\DeclareMathSymbol{\squarebotblack}{\mathord}{SYMB4}{"CE}
\DeclareMathSymbol{\mdwhtcircle}{\mathord}{ARR3}{"7D}
\DeclareMathSymbol{\mdwhtsquare}{\mathord}{ARR3}{"9A}

\definecolor{darkviolet}{rgb}{0.58, 0.0, 0.83}
\definecolor{shamrockgreen}{rgb}{0.0, 0.62, 0.38}
\definecolor{deepskyblue}{rgb}{0.0, 0.75, 1.0}
\definecolor{amber}{rgb}{1.0, 0.49, 0.0}
\definecolor{red}{rgb}{1.0, 0.0, 0.0}

\newcommand\solid[1][1.0cm]{\rule[0.4ex]{#1}{2.8pt}}

\newcommand{\diff}{\mathrm{d}}

 \title[Turbulence in pipe flow]{One-point statistics for turbulent pipe flow \\ up to $\Rey_{\tau} \approx 6000$}

\author{Sergio Pirozzoli\aff{1}
  \corresp{\email{sergio.pirozzoli@uniroma1.it}},
  Joshua Romero\aff{2},
  Massimiliano Fatica\aff{2},
  Roberto Verzicco\aff{3,4},
 \and Paolo Orlandi\aff{1}}

\affiliation{
\aff{1} Dipartimento di Ingegneria Meccanica e Aerospaziale, Sapienza Universit\`a di Roma, Via Eudossiana 18, 00184 Roma, Italy
\aff{2} NVIDIA Corporation, 2701 San Tomas Expressway, Santa Clara, CA 95050, USA
\aff{3} Dipartimento di Ingegneria Industriale, Universit\`a di Roma TorVergata, Via del Politecnico 1, 00133 Roma, Italy
\aff{4} Physics of Fluid Group, University of Twente, P.O. Box 217, 7500 AE Enschede, The Netherlands
}

\date{\today}

\begin{document}

\maketitle

\begin{abstract}
We study turbulent flows in a smooth straight pipe of circular cross--section 
up to $\Rey_{\tau} \approx 6000$ using direct--numerical-simulation (DNS)
of the Navier--Stokes equations.
The DNS results highlight systematic deviations from Prandtl friction law, 
amounting to about $2\%$, which would extrapolate to about $4\%$ at extreme Reynolds numbers. 
Data fitting of the DNS friction coefficient yields 
an estimated von K\'arm\'an constant $k \approx 0.387$, which nicely fits the mean velocity profile,
and which supports universality of canonical wall-bounded flows.
The same constant also applies to the pipe centerline velocity, thus providing support for the claim that the 
asymptotic state of pipe flow at extreme Reynolds numbers should be plug flow.
At the Reynolds numbers under scrutiny, no evidence for saturation of the logarithmic growth 
of the inner peak of the axial velocity variance is found.
Although no outer peak of the velocity variance directly emerges in our DNS, 
we provide strong evidence that it should appear at $\Rey_{\tau} \gtrsim 10^4$,
as a result of turbulence production exceeding dissipation over a large part of the outer wall layer,
thus invalidating the classical equilibrium hypothesis.
\end{abstract}

\section{Introduction}\label{sec:intro}

Turbulent flow in circular pipes has always attracted the interest
of scientists, owing to its prominent importance in the engineering practice 
and because of the beautiful simplicity of the setup. In this respect, the pioneering flow
visualizations of \citet{reynolds_1883} may be regarded as a milestone for the
understanding of turbulent and transitional flows. 
The most extensive experimental measurements of high-Reynolds-number pipe flows have been
carried out in modern times in the Princeton Superpipe pressurized facility~\citep{zagarola_98,mckeon_05,hultmark_10}. 
Those investigations have allowed scientists to measure the main flow features 
as friction and mean velocity profiles with high precision, 
and they currently constitute the most comprehensive database for the study of
pipe turbulence. However, even the use of specialized micro-fabricated hot-wire probes 
could not provide fully reliable information about the viscous and buffer layers
at high Reynolds numbers~\citep{hultmark_12}.
Additional experimental studies of pipe turbulence have been carried out in the 
high-Reynolds-number actual flow facility (Hi-Reff), a water tunnel 
with relatively large diameter, which allows for accurate estimation 
of friction~\citep{furuichi_15,furuichi_18}.
Recently, the CICLoPE facility of the University of Bologna~\citep{fiorini_17, willert_17}
has been set up, whose large diameter (about 1m) offers a 
well-established turbulent flow with relatively large viscous scales, thus granting 
higher spatial resolution.
Flows in different facilities seem to have sensibly different properties in terms of
friction and mean velocity profiles, which we will comment on.


Numerical simulation of pipe turbulence flow has received less interest than other
canonical flows, the plane channel in particular, because of additional difficulties 
involved with discrete solution of the Navier--Stokes equations in cylindrical coordinates,
with special reference to treatment of the geometrical singularity at the pipe axis.
Early numerical simulations of turbulent pipe flow were carried out by~\citet{eggels_94},
at friction Reynolds number $\Rey_{\tau}=180$ ($\Rey_{\tau} = u_{\tau} R / \nu$,
with $u_{\tau} = (\tau_w/\rho)^{1/2}$ the friction velocity, $R$ the pipe radius, 
and $\nu$ the fluid kinematic viscosity).
Effects of drag reduction associated with pipe rotation were later studied by~\citet{orlandi_97}.
Higher Reynolds numbers (up to $\Rey_{\tau} \approx 1140$) were reached by \citet{wu_08},
which first allowed to observe a near logarithmic layer in the mean velocity profile.
Flow visualizations and two-point correlation statistics pointed to 
the existence of high-speed wavy structures in the pipe core region
which are elongated in the axial direction, and whose
streamwise and azimuthal dimensions do not change substantially with the Reynolds number,
when normalized in outer units.
Further follow-up DNS studies have been carried out by \citet{elkhoury_13, chin_14, ahn_13}. 
At present, the highest Reynolds number in pipe flow ($\Rey_{\tau} \approx 3000$)
has been reached in the study of \citet{ahn_15}. Although no sizeable logarithmic layer
is present yet at those conditions, some effects associated with significant scale
separation between inner- and outer-scale turbulence were observed,
as the presence of a $k^{-1}$ ($k$ being the wavenumber in any wall-parallel direction) 
power-law ranges in the velocity spectra.

Despite inherent limitations in the Reynolds numbers which can be attained,
DNS has the advantage over experiments of yielding immediate access to the near-wall region, 
and of allowing scientists to measure some flow properties, e.g. the turbulence dissipation rate, which 
can hardly be measured in experiments. Hence, it is generally claimed that
DNS data at increasing Reynolds numbers are needed to prove or disprove 
theoretical claims related to departure (or not) of the statistical properties
of wall-bounded turbulence from the universal wall scaling~\citep{cantwell_19, monkewitz_21, chen_21}.
In this paper we thus present DNS data of turbulent flow in a smooth circular
pipe at $\Rey_{\tau} \approx 6000$, 
which is two times higher that the previous state of art.
Relying on the DNS data, we revisit current theoretical inferences and 
discuss implications about possible trends in the extreme Reynolds number regime.

\section{The numerical dataset}\label{sec:numerics}

The code used for DNS is the spin-off of an existing solver previously used to
study Rayleigh-B\'enard convection in cylindrical containers at
extreme Rayleigh numbers~\citep{stevens_13}.
That code is in turn the evolution of the solver originally developed 
by \citet{verzicco_96}, and used for DNS of pipe flow by \citet{orlandi_97}.
A second-order finite-difference discretization of the 
incompressible Navier-Stokes equations in cylindrical coordinates is used, 
based on the classical marker-and-cell method~\citep{harlow_65}, with staggered arrangement of the flow variables
to remove odd-even decoupling phenomena and guarantee
discrete conservation of the total kinetic energy in the inviscid flow limit.
Uniform volumetric forcing is applied to the axial momentum equation 
to maintain constant mass flow rate in time.
The Poisson equation resulting from enforcement of the divergence-free condition is
efficiently solved by double trigonometric expansion in the periodic axial
and azimuthal directions, and inversion of tridiagonal matrices in the radial direction~\citep{kim_85}.
An extensive series of previous studies about wall-bounded flows from this group proved
that second-order finite-difference discretization yields in practical cases of wall-bounded turbulence results 
which are by no means inferior in quality to those of pseudo-spectral methods~\citep[e.g.][]{pirozzoli_16,moin_16}.
A crucial issue is the proper treatment of the polar singularity at the pipe axis.
A detailed description of the subject is reported in \citet{verzicco_96}, but basically,
the radial velocity $u_r$ in the governing equations is replaced by $q_r = r u_r$
($r$ is the radial space coordinate), which by construction vanishes at the axis. 
The governing equations are advanced in time by means of a hybrid third-order low-storage
Runge-Kutta algorithm, whereby the diffusive terms are handled implicitly, and convective terms in the axial and radial direction explicitly.
An important issue in this respect is the convective time step limitation in the azimuthal direction,
due to intrinsic shrinking of the cells size toward the pipe axis. To alleviate this limitation
we rely on implicit treatment of the convective terms in the azimuthal direction~\citep{akselvoll_96,wu_08}, 
which enables marching in time with similar time step as in planar domains flow in practical computations.
In order to minimize numerical errors associated with implicit time stepping,
in the present code explicit and explicit discretizations of the azimuthal convective terms are linearly blended 
with the radial coordinate, in such a way that near the pipe wall the treatment is fully
explicit, and near the pipe axis it is fully implicit.
The code was adapted to run on clusters of graphic accelerators (GPUs), using a combination of 
CUDA Fortran and OpenACC directives, and relying on the CUFFT libraries for efficient 
execution of FFTs~\citep{fatica_14}. The DNS were carried out on the Marconi-100 machine
based at CINECA (Italy), relying on NVIDIA Volta V100 graphic cards. 
Specifically, 1024 GPUs were used for DNS-F.

\begin{table}
 \centering
\begin{tabular*}{1.\textwidth}{@{\extracolsep{\fill}}lcccccccc}
 Dataset & $L_z/R$ & Mesh ($N_\theta \times N_r \times N_z$) & $\Rey_b$ & $\lambda$ & $\Rey_{\tau}$ & $T/\tau_t$ & Line style \\
 \hline
 DNS-A   & 15  & $256  \times 67  \times  256$ &  $5300$  & $0.03700$ & $180.3$ & 204.0 & \color{darkviolet}\solid \\
 DNS-B   & 15  & $768  \times 140 \times  768$ & $17000$  & $0.02716$ & $495.3$ &  87.4 & \color{shamrockgreen}\solid \\
 DNS-C   & 15  & $1792 \times 270 \times 1792$ & $44000$  & $0.02136$ & $1136.6$ & 25.9 & \color{deepskyblue}\solid \\
 DNS-C-SH   & 7.5 & $1792 \times 270 \times 986$ & $44000$  & $0.02164$ & $1144.2$ & 31.1 & NA \\
 DNS-C-LO   & 30  & $1792 \times 270 \times 3944$ & $44000$  & $0.02128$ & $1134.6$ & 24.5 & NA \\
 DNS-C-FT   & 15  & $3944 \times 270 \times 1792$ & $44000$  & $0.02114$ & $1131.0$ & 31.3 & NA \\
 DNS-C-FR   & 15  & $1792 \times 540 \times 1792$ & $44000$  & $0.02132$ & $1135.7$ & 28.6 & NA \\
 DNS-C-FZ   & 15  & $1792 \times 270 \times 3944$ & $44000$  & $0.02132$ & $1135.7$ & 15.5 & NA \\
 DNS-D   & 15  & $3072 \times 399 \times 3072$ & $82500$  & $0.01836$ & $1976.0$ & 22.4 & \color{amber}\solid \\
 DNS-E   & 15  & $4608 \times 540 \times 4608$ & $133000$ & $0.01659$ & $3028.1$ & 16.6 & \color{red}\solid \\
 DNS-F   & 15  & $9216 \times 910 \times 9216$ & $285000$ & $0.01428$ & $6019.4$ & 8.32 & \solid \\
 \hline
\end{tabular*}
\caption{Flow parameters for DNS of pipe flow.
$R$ is the pipe radius, $L_z$ is the pipe axial length,
$N_{\theta}$, $N_r$ and $N_z$ are the number of grid points in the 
azimuthal, radial and axial directions, respectively, 
$\Rey_b = 2 R u_b / \nu$ is the bulk Reynolds number,
$\lambda = 8 \tau_w / (\rho u_b^2)$ is the friction factor,
$\Rey_{\tau} = u_{\tau} R / \nu$ is the friction Reynolds number,
$T$ is the time interval used to collect the flow statistics,
and $\tau_t = R/u_{\tau}$ is the eddy turnover time.
}
\label{tab:runs}
\end{table}

\begin{table}
 \centering
\begin{tabular*}{1.\textwidth}{@{\extracolsep{\fill}}lccccc}
 Dataset  & $\lambda$ & $U_{_{CL}}^+$ & $<u_z^2>^+_{_{IP}}$ & $y^+_{_{IP}}$ & $\epsilon^+_{_{11w}}$ \\
 \hline
 DNS-A    & $ 0.03700  \pm 0.15\%  $ & $ 19.30 \pm 0.087\% $ & $ 7.129   \pm 0.26\% $ & $ 14.95 \pm 0.24\%  $ & $ 0.1168   \pm 0.47\%    $ \\
 DNS-B    & $ 0.02716  \pm 0.074\% $ & $ 21.81 \pm 0.17\%  $ & $ 7.352   \pm 0.17\% $ & $ 14.28 \pm 0.010\% $ & $ 0.1506   \pm 0.21\%    $ \\
 DNS-C    & $ 0.02136  \pm 0.13\%  $ & $ 24.07 \pm 0.18\%  $ & $ 7.995   \pm 0.29\% $ & $ 14.66 \pm 0.073\% $ & $ 0.1697   \pm 0.37\%    $ \\
 DNS-C-SH & $ 0.02164  \pm 0.14\%  $ & $ 24.09 \pm 0.20\%  $ & $ 8.071   \pm 0.44\% $ & $ 14.37 \pm 0.11\%  $ & $ 0.1952   \pm 0.54\%    $ \\
 DNS-C-LO & $ 0.02128  \pm 0.16\%  $ & $ 24.17 \pm 0.11\%  $ & $ 7.965   \pm 0.29\% $ & $ 14.62 \pm 0.058\% $ & $ 0.1704   \pm 0.40\%    $ \\
 DNS-C-FT & $ 0.02114  \pm 0.12\%  $ & $ 24.28 \pm 0.14\%  $ & $ 7.948   \pm 0.27\% $ & $ 14.66 \pm 0.078\% $ & $ 0.1691   \pm 0.34\%    $ \\
 DNS-C-FR & $ 0.02132  \pm 0.25\%  $ & $ 24.10 \pm 0.12\%  $ & $ 7.886   \pm 0.31\% $ & $ 14.41 \pm 0.096\% $ & $ 0.1741   \pm 0.60\%    $ \\
 DNS-C-FZ & $ 0.02132  \pm 0.21\%  $ & $ 24.07 \pm 0.26\%  $ & $ 8.168   \pm 0.38\% $ & $ 14.89 \pm 0.14\%  $ & $ 0.1727   \pm 0.44\%    $ \\
 DNS-D    & $ 0.01839  \pm 0.25\%  $ & $ 25.56 \pm 0.34\%  $ & $ 8.397   \pm 0.43\% $ & $ 14.79 \pm 0.098\% $ & $ 0.1822   \pm 0.57\%    $ \\
 DNS-E    & $ 0.01658  \pm 0.26\%  $ & $ 26.47 \pm 0.27\%  $ & $ 8.681   \pm 0.69\% $ & $ 14.87 \pm 0.13\%  $ & $ 0.1903   \pm 0.93\%    $ \\
 DNS-F    & $ 0.01428  \pm 0.36\%  $ & $ 28.05 \pm 0.35\%  $ & $ 9.108   \pm 0.72\% $ & $ 15.14 \pm 0.20\%  $ & $ 0.1993   \pm 1.10\%    $ \\
 \hline
\end{tabular*}
\caption{Uncertainty estimation study: mean values of representative quantities and standard deviation of their estimates. $\lambda$ is the friction factor, $U_{_{CL}}^+$ is the mean pipe centerline velocity, $<u_z^2>^+_{_{IP}}$ is the peak axial velocity variance and $y^+_{_{IP}}$ is its distance from the wall, and $\epsilon^+_{_{11w}}$ is the 
dissipation rate of $<u_z^2>$ at the wall.}
\label{tab:UQ}
\end{table}

\begin{table}
 \centering
\begin{tabular*}{1.\textwidth}{@{\extracolsep{\fill}}lccc}
 Source & Type & $\Rey_{\tau}$ range & Symbols \\
 \hline
 \citet{wu_08}       & DNS   & 180, 1140   & \color{shamrockgreen}$\blacksquare$ \\
 \citet{elkhoury_13} & DNS   & 180-1000    & \color{amber}$\pentagonblack$ \\
 \citet{chin_14}     & DNS   & 180-2000    & \color{deepskyblue}$\blacktriangle$ \\
 \citet{ahn_13}, \citet{ahn_15} & DNS      & 180-3000 & \color{darkviolet}$\mdlgblkdiamond$ \\
 \citet{durst_95}    & EXP   & 250         & \color{darkviolet}$\mdlgwhtdiamond$ \\
 \citet{swanson_02}  & EXP   & 170-1500    & \color{amber}$\mdwhtsquare$ \\
 \citet{fiorini_17}  & EXP   & 3000-35000  & \color{shamrockgreen}$\circletophalfblack$ \\
 \citet{willert_17}  & EXP   & 5400-40000  & \color{shamrockgreen}$\circlerighthalfblack$ \\
 \citet{nagib_17}    & EXP   & 8000-40000  & \color{shamrockgreen}$\circlebottomhalfblack$ \\
 \citet{mckeon_05}   & EXP   & 1800-32900  & \color{gray}$\squaretopblack$ \\
 \citet{hultmark_12} & EXP   & 2000-20000  & \color{gray}$\squarebotblack$ \\
 \citet{furuichi_15}, \citet{furuichi_18} & EXP &  200-53000  & \color{red}$\bigtriangledown$ \\
 \citet{flack_13} & EXP (channel) &  1000-6000  & \color{amber}$\mdwhtcircle$ \\
 \citet{lee_15} & DNS (channel) &  180-5200  & \color{red}$\mdwhtcircle$ \\
 \hline
\end{tabular*}
\caption{List of other references for data used in the paper}
\label{tab:refs}
\end{table}

Numerical simulations are carried out with periodic 
boundary conditions in the axial ($z$) and azimuthal ($\theta$) directions.
The velocity field is then controlled by two parameters, namely the bulk
Reynolds number ($\Rey_b = 2 R u_b / \nu$, with $R$ the pipe radius, $u_b$ the fluid bulk velocity,
and $\nu$ its kinematic viscosity), and the relative
pipe length, $L_z/R$. 
A list of the main simulations that we have carried out is given in table~\ref{tab:runs}.
The mesh resolution is designed based on well-established criteria in the wall turbulence community.
In particular, the collocation points are distributed in the wall-normal direction so that about thirty points are placed within $y^+ \le 40$ ($y=R-r$ is the wall distance, and 
the + superscript is used to denote normalization with respect to $u_{\tau}$ and $\nu$), with the first grid point at $y^+ \approx 0.05$.
The mesh is progressively stretched in the outer wall layer in such a way that the mesh spacing is proportional to the local Kolmogorov 
length scale, which there varies as $\eta^+ \approx 0.8 \, {y^+}^{1/4}$~\citep{jimenez_18}, and the radial spacing at the 
pipe axis is $\Delta y^+ \approx 8.8$. 
Additional details are provided in a specifically focused publication~\citep{pirozzoli_21}.
Regarding the axial and azimuthal directions, finite-difference simulations of wall-bounded flows yield grid-independent results as long as $\Delta x^+ \approx 10$, $R^+ \Delta \theta \approx 4.5$~\citep{pirozzoli_16},
hence the associated number of grid points scales as 
$N_z \approx L_z / R \times \Rey_{\tau} / 10$, $N_{\theta} \sim 2 \pi \times \Rey_{\tau} / 4.5$.
All DNS have been carried out at CFL number close to unity, based on the 
radial convective time step limitation. The CFL number along the axial direction is typically smaller
by a factor two. The time step expressed in wall units ($\nu / u_{\tau}^2$) ranges from 
$\Delta t^+ = 0.55$ in DNS-A to $\Delta t^+ = 0.15$ in DNS-F.
According to the established practice~\citep{hoyas_06, lee_15, ahn_15}, the time intervals used 
to collect the flow statistics are reported in terms of eddy-turnover times, $\tau_t = R/u_{\tau}$. 
For reference, the time window used to collect the flow statistics in DNS-F amounts to about 
13.1 flow-through times ($L_z/u_b$ time units).

The sampling errors for some key properties discussed in this paper have been estimated using the method of \citet{russo_17}, 
based on extension of the classical batch means approach. 
The results of the uncertainty estimation analysis are listed in table~\ref{tab:UQ}, 
where we provide expected values and associated standard deviation for the friction factor ($f$),
mean centerline velocity ($U_{_{CL}}$), peak axial velocity variance and its position ($\left<{u_z^2}\right>_{_{IP}}$ and
$y_{_{IP}}$, respectively), and dissipation rate of streamwise velocity variance ($\epsilon_{_{11w}}$).
Here and elsewhere, capital letters are used to denote flow properties averaged in the 
homogeneous spatial directions and in time, brackets denote the averaging operator, 
and lower-case letters to denote fluctuations from the mean.
We find that the sampling error is generally quite limited, being larger in the largest DNS, 
which have been run for shorter time. In particular, in DNS-F the expected sampling error in friction, 
centerline velocity and peak velocity variance is about $0.5 \%$, whereas it is about $1\%$ for the wall dissipation.
Additional tests aimed at establishing the effect of axial domain length and grid size
have been carried out for the DNS-C flow case, whose results are also reported in table~\ref{tab:UQ}.
We find that doubling the pipe length yields a change in the basic flow properties of about $0.2-0.3\%$,
whereas halving it yields changes of about $1\%$ in friction and peak velocity variance, 
and up to $10\%$ in the wall dissipation. Hence, consistent with previous studies~\citep{chin_10}, we believe that
the selected pipe length ($L_z/R=15$) is representative of an infinitely long pipe, at least for 
the purposes of the present study. In order to quantify uncertainties associated with numerical 
discretization, additional simulations have been carried out by doubling the grid points 
in the azimuthal, radial and axial directions, respectively.
Based on the data reported in the table, after discarding the short pipe case, we can thus quantify the uncertainty due 
to numerical discretization and limited pipe length to be about $0.3\%$ for the friction coefficient and pipe centerline velocity,
$0.6\%$ for the peak velocity variance, and $0.9\%$ for the wall dissipation.

\section{Results} \label{sec:results}

\begin{figure}
 \centerline{
 \includegraphics[width=15.0cm]{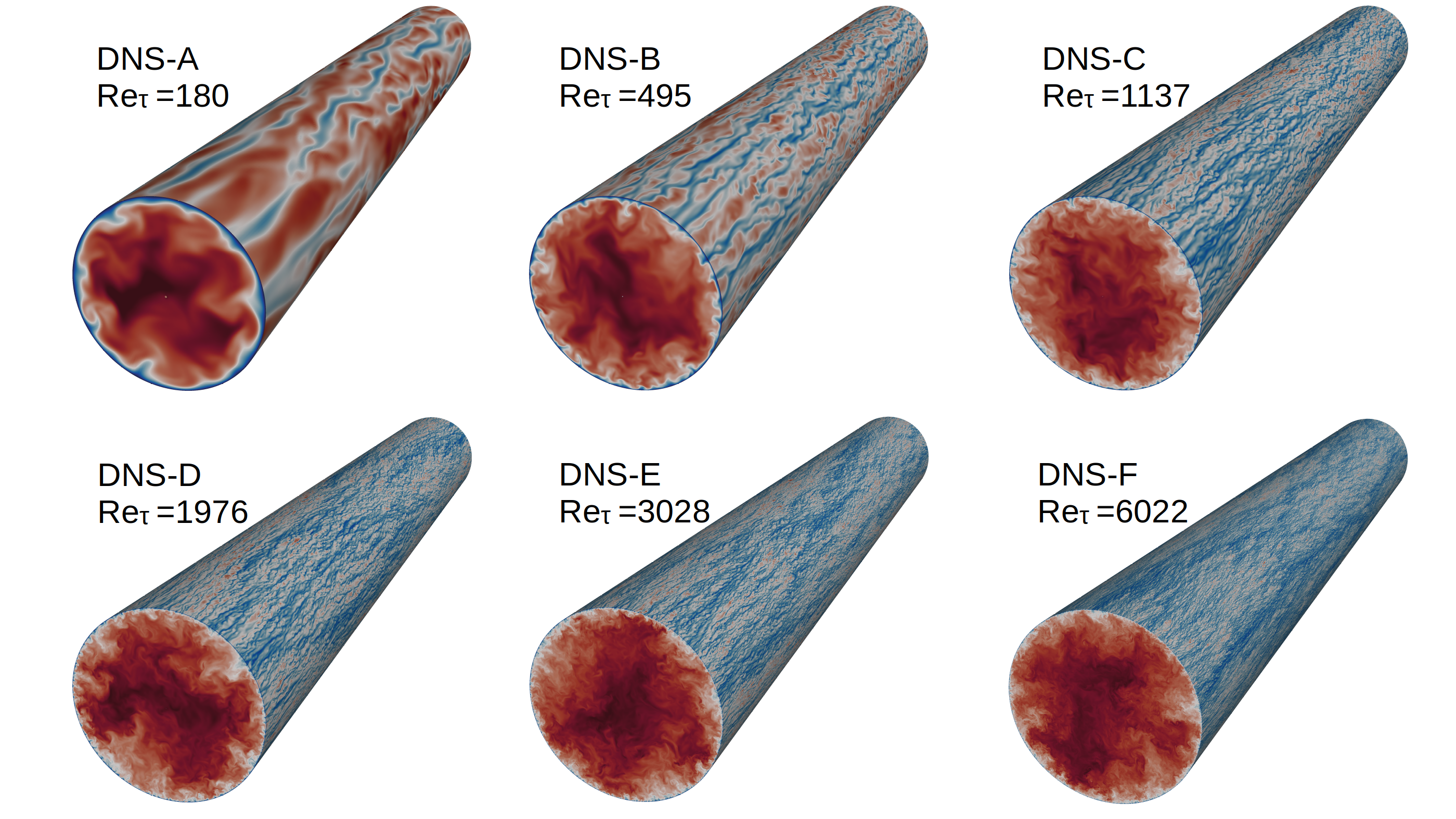}
 }
\caption{
Instantaneous streamwise velocity contours (colour scale from blue to red) in turbulent pipe flow as obtained from DNS. 
Contours are shown on a cross-stream plane and on a near-wall cylindrical shell ($y^+ \approx 15$). 
}
\label{fig:3D}
\end{figure}

Qualitative information about the structure of the flow field is provided by instantaneous 
perspective views of the axial velocity field, provided in figure~\ref{fig:3D}.
Although finer-scale details are visible at the higher $\Rey$,
the flow in the cross-stream planes is always characterized 
by a limited number of bulges distributed along the azimuthal direction, which closely recall the POD modes
identified by \citet{hellstrom_14}, and which correspond to alternating 
intrusions of high-speed fluid from the pipe core and ejections of low-speed fluid from the wall.
Streaks are visible in the near-wall cylindrical shells, whose organization has clear association
with the cross-stream pattern. Specifically, regardless of the Reynolds number, $R$-sized
low-streaks are observed in association with large-scale ejections, whereas
$R$-sized high-speed streaks occur in the presence of large-scale inrush from the core flow.
At the same time, smaller streaks scaling in wall units appear, corresponding to buffer-layer
ejections/sweeps. Hence, organization of the flow on at least two length scales is apparent here,
whose separation increases with $\Rey_{\tau}$.


\begin{figure}
\centerline{
(a)~\includegraphics[width=7.0cm]{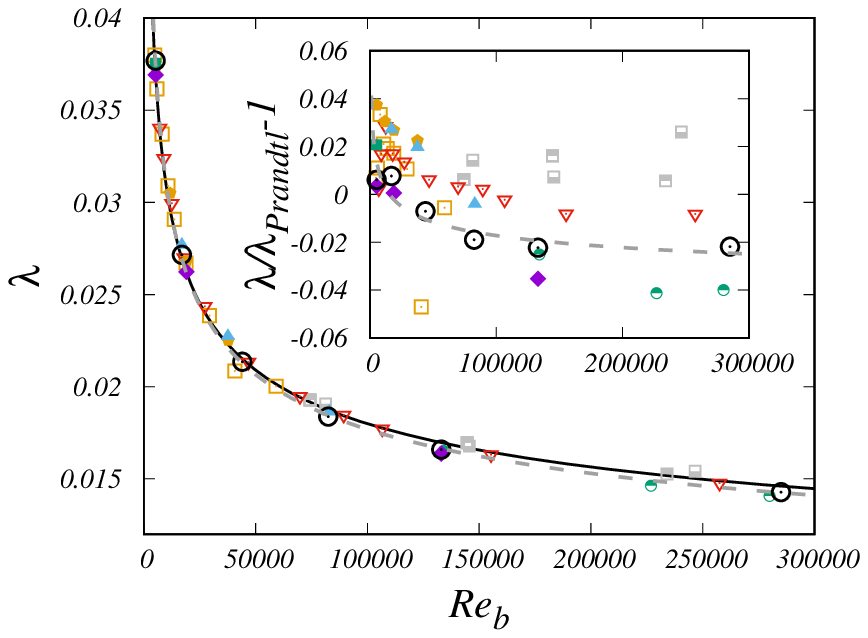}~(b)~\includegraphics[width=7.0cm]{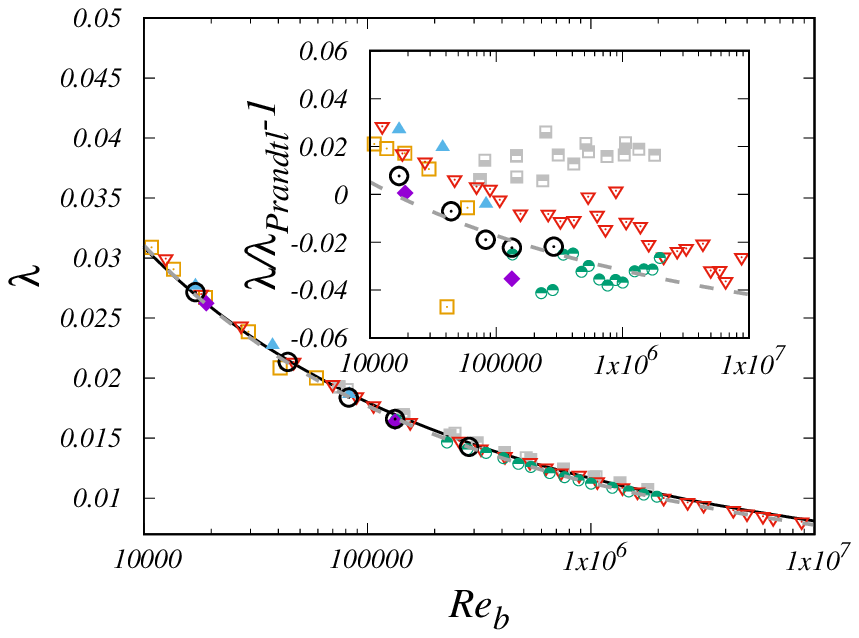}
}
\caption{Friction factor as a function of bulk Reynolds number, in linear (a) and in semi-logarithmic (b) scale. 
Circles denote present DNS data, other symbols are defined in table~\ref{tab:refs}. The solid line corresponds to the
classical Prandtl friction law as given in equation~\eqref{eq:prandtl}, whereas the dashed grey line corresponds to
a fit of the DNS data. Relative deviations with respect to the Prandtl friction law are shown in the insets.
}
\label{fig:cf}
\end{figure}

Mean friction is obviously a parameter of paramount importance as it is related to power expenditure to sustain the flow.
In figure~\ref{fig:cf}, we show the friction factor, namely
\begin{equation}
\lambda = \frac {8 \tau_w}{\rho u_b^2}. \label{eq:Cf}
\end{equation}
A correlation generally used for smooth pipes is the Prandtl friction law,
\begin{equation}
1/{\lambda}^{1/2} = A \log_{10} (\Rey_{b} {\lambda}^{1/2}) - B, \label{eq:prandtl}
\end{equation}
where $A= \log 10 / (2 k \sqrt{2})$, with $k$ the von K\'arm\'an log-law constant.
The standard values $A=2.0$, $B=0.8$, were derived by fitting the experimental data of \citet{nikuradse_33}.
Reynolds-number-dependent corrections to the standard friction law were
introduced by \citet{mckeon_05} in order to improve the fitting of Superpipe data.
Figure~\ref{fig:cf} shows overall agreement of all DNS and experimental data with Prandtl law. 
However, closer scrutiny (see the figure insets) highlights some scatter.
Regarding DNS, all datasets overshoot Prandtl law at low Reynolds number, although to a quite different extent.
In fact, the data of \citet{wu_08}, \citet{elkhoury_13}, \citet{chin_14} exceed the theoretical values by up to $4\%$,
whereas our data tend to be much more consistent with those of \citet{ahn_15}. We believe that 
this difference may be related to different grid resolution in the azimuthal direction,
which was $R^+ \Delta \theta = 7-8$ in those previous studies, and $4-5$ in our DNS.
Our data in fact show minimal overshoot at low Reynolds number, and 
consistent undershoot from Prandtl law by about $2\%$.
Regarding experiments, Superpipe data typically tend to lie above the theoretical
curve by about $2\%$, whereas the CICLoPE and Hi-Reff data tend to fall short of it.
Although the range of data overlap is not extensive,
it appears that DNS data tend to be more consistent with the CICLoPE and Hi-Reff data than with other datasets. 
Fitting the current DNS data 
with a functional relationship as (\ref{eq:prandtl}), yields $A \approx 2.102$, $B \approx 1.148$,
with an inferred value of the von K\'arm\'an constant of $k = 0.387 \pm 0.004$, 
with uncertainty estimates based on $95\%$ confidence bounds from the curve-fitting procedure.
This value is extremely close to that suggested by \citet{furuichi_18}, who reported $k=0.386$
as an average value over a very wide range of Reynolds numbers, and also very close to values reported
in boundary layers~\citep{chauhan_09} and channels~\citep{lee_15}.
If this trend is extrapolated, deviations of about $4\%$ from the standard Prandtl law would result at $\Rey_b = 10^7$.


\begin{figure}
 \centerline{
 (a)~\includegraphics[width=7.0cm]{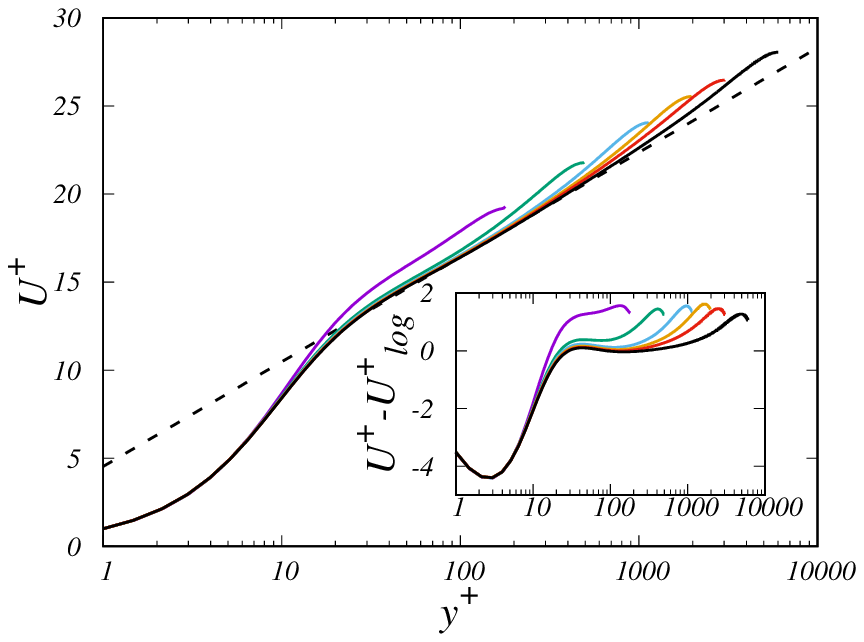}~(b)~\includegraphics[width=7.0cm]{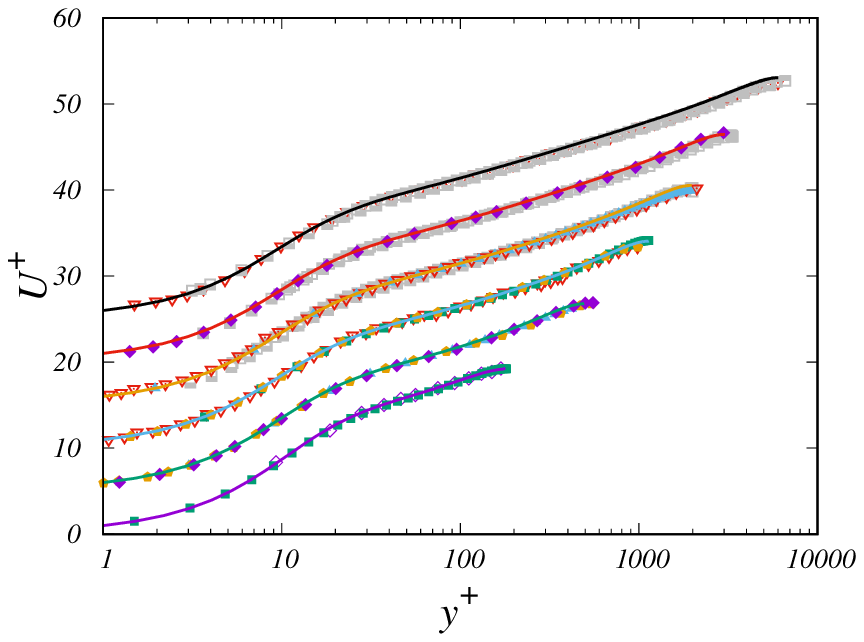}}
\caption{Inner-scaled mean velocity profiles obtained with our DNS (a), and compared with previous DNS and experiments (b). 
Deviations from the assumed logarithmic wall law, $U_{log}^+ = \log y^+ / 0.387 + 4.53$, are highlighted in the inset of panel (a).
For greater clarity, profiles in panel (b) are offset in the vertical direction by five wall units steps. 
Lines denote present DNS data, with color code as in table~\ref{tab:runs}, and symbols denote data from other authors, as in table~\ref{tab:refs}.}
\label{fig:uplus}
\end{figure}

The mean velocity profile in turbulent pipes 
has received extensive attention from theoretical studies, much of the early debate being focused on
whether a log law or a power law better fits the experimental results~\citep{barenblatt_97}, 
mainly carried out in the Superpipe facility~\citep{zagarola_98,mckeon_05}. Recent studies have 
highlighted the need for corrections to the baseline log law in order to accurately describe the velocity profile throughout the log layer
into the core part of the flow~\citep{luchini_17,cantwell_19,monkewitz_21}. 
In figure~\ref{fig:uplus}, we show the series of velocity profiles computed with the present DNS, 
compared with previous DNS and experimental data. Overall, good agreement is observed across various sources as far as the inner and the 
overlap regions are concerned, with data gradually approaching a logarithmic distribution,
here identified by visual fitting as $U^+ = 1/k \log y^+ + 4.53$, using the value of $k=0.387$ determined from friction data.
This is quite close to estimates based on direct fitting of the mean velocity profile in pipe flow~\citep{marusic_13},
which yielded $U^+ = 1/0.391 \log y^+ + 4.34$.
The DNS velocity profiles for $\Rey_{\tau} \ge 10^3$ follow this distribution with deviations of no more than $0.1$ wall units
from $y^+ \approx 30$ to $y/R \approx 0.15$, whence the core region develops. 
Differences with respect to previous DNSs are concentrated in the core region,
which seemingly stronger wake in some datasets, including our own, \citet{wu_08} and \citet{ahn_13}, 
and weaker in others~\citep{elkhoury_13,chin_14}, reflecting previously noted differences in the friction coefficient.
Especially satisfactory is the excellent agreement between our DNS-E velocity profile and 
the data of \citet{ahn_15} at $\Rey_{\tau} \approx 3000$. 
Comparison of our DNS dataset with experimental data also shows overall good
agreement, although some differences are quite clear in the core region, in which Superpipe experiments 
consistently yield lower $U^+$, which translates into lower friction.

\begin{figure}
 \centerline{
 (a)~\includegraphics[height=5.0cm]{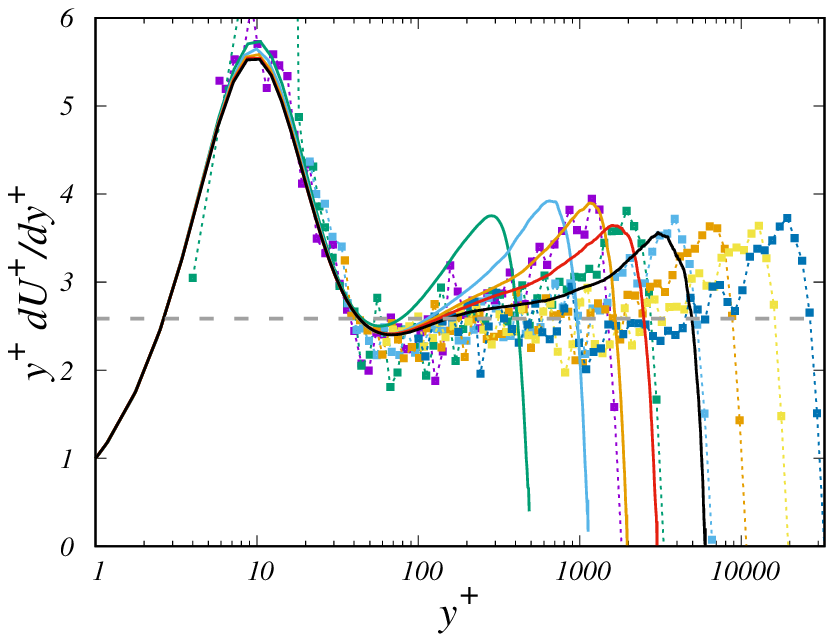}~(b)~\includegraphics[height=5.0cm]{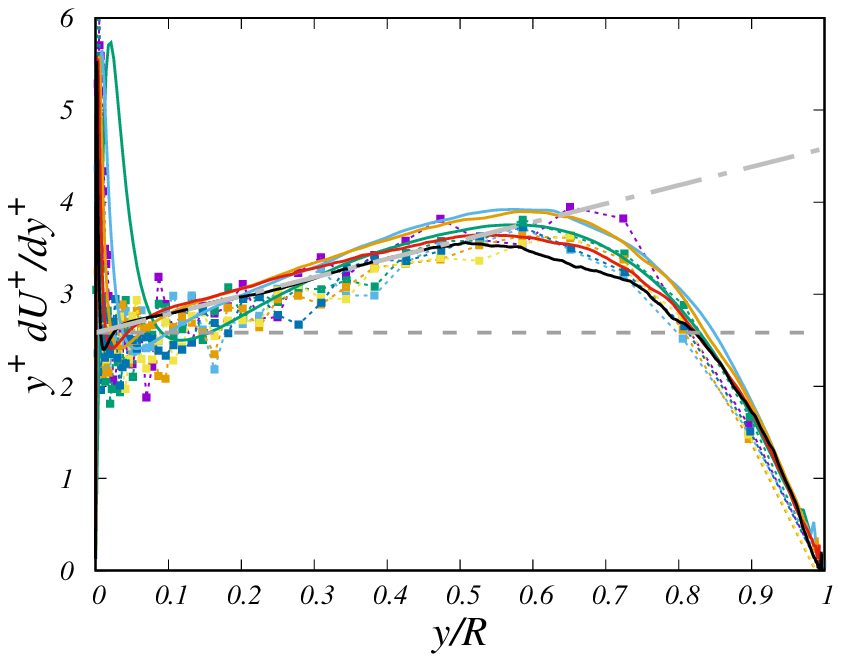}}
\caption{Log-law diagnostic function as defined in equation~\eqref{eq:diag}, expressed as a 
function of inner-scaled (a) and outer-scaled (b) wall distance.
The dashed horizontal line denotes the inverse K\'arm\'an constant, $1/0.387$, 
and the dash-dotted lines in panel (b) denotes the linear fit \eqref{eq:jimfit}, with $k=0.387$, $\alpha = 2.0$, $\beta=0$.
Lines denote present DNS data, with color code as in table~\ref{tab:runs}, and symbols denote Superpipe data~\citep{mckeon_05} at
$\Rey_{\tau} = 1825, 3328, 6617, 10914, 19119, 32870$.}
\label{fig:diag}
\end{figure}

More refined information on the behaviour of the mean velocity profile can be gained from inspection
of the log-law diagnostic function
\begin{equation}
\Xi = y^+ \, \diff {U}^+ / \diff y^+ , \label{eq:diag}
\end{equation}
which is shown in figure~\ref{fig:diag}, and whose constancy would imply the presence of a genuine logarithmic layer in the mean velocity profile.
The figure supports universality of the inner-scaled axial velocity for $\Rey_{\tau} \gtrsim 10^3$, 
up to $y^+ \approx 100$, where $\Xi$ attains a
minimum, and the presence of an outer maximum at $y/R \approx 0.6$.
Between these two sites the distribution is roughly linear, as can be better appreciated 
in panel (b), with nearly constant slope when expressed in outer coordinates.
Approximate linear variation of the diagnostic function in channel flow was observed by \citet{jimenez_07}, 
who, based on refined overlap arguments expressed by \citet{afzal_73}, proposed the following fit
\begin{equation}
\Xi = \frac 1k + \frac {\beta}{\Rey_{\tau}} + \alpha \frac yR , \label{eq:jimfit}
\end{equation}
where $\alpha$, $\beta$ are adjustable constants, and $k$ is the von K\'arm\'an constant.
Here we find that the set of constants $k=0.387$, $\alpha=2.0$, $\beta=0$,
yields overall good approximation of the pipe DNS data.
The consequence is that a genuine logarithmic layer would only be attained at infinite Reynolds number.
In this respect, Superpipe data seem to suggest the formation of a plateau at $\Rey_{\tau} \gtrsim 10^4$, 
although the scatter of points is quite significant.
Hence, DNS at higher Reynolds number would be most welcome to confirm or refute our findings,
and possibly determine more accurate values of the extended log-law constants in \eqref{eq:jimfit}.

\begin{figure}
 \centerline{
 (a)~\includegraphics[width=7.0cm]{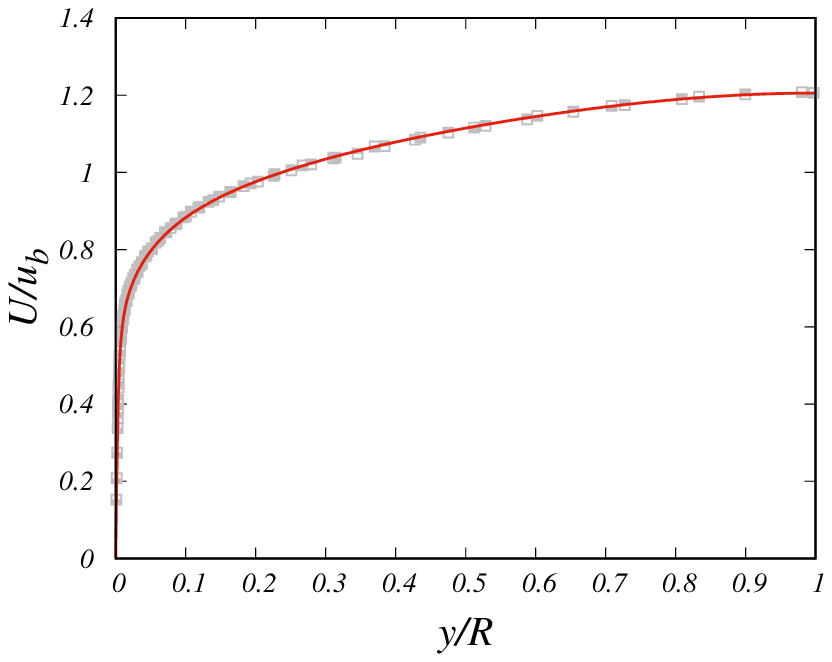}~(b)~\includegraphics[width=7.0cm]{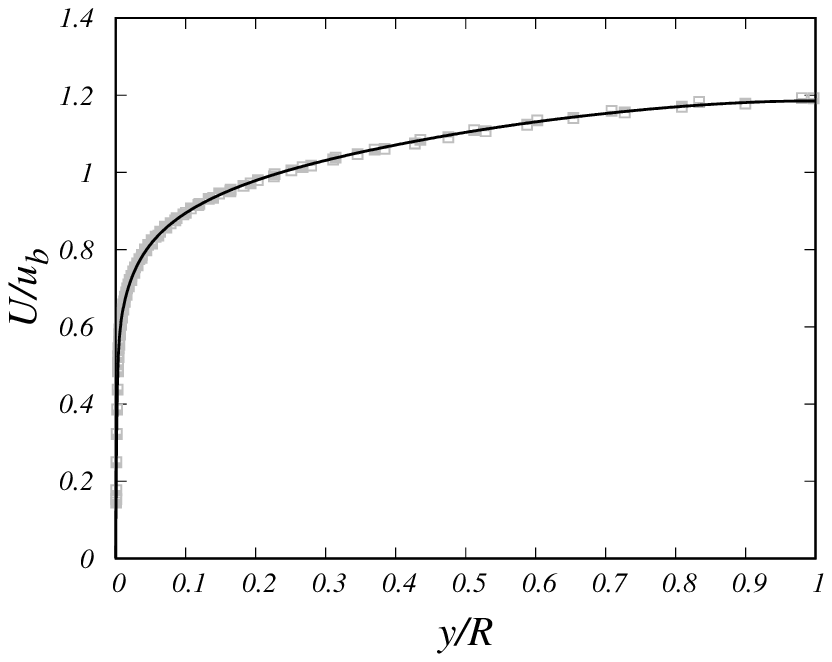}}
\caption{Mean velocity profiles in outer scaling. Data of flow case DNS-E (left) are compared with Superpipe
data at $\Rey_{\tau}=3328$ and $\Rey_{\tau}=3334$, and data of flow case DNS-F (right) with Superpipe data at $\Rey_{\tau}=5411$ and $\Rey_{\tau}=6617$.}
\label{fig:uouter}
\end{figure}

Comparison with Superpipe data is presented in outer units in figure~\ref{fig:uouter}, limited
to the higher $\Rey_{\tau}$ cases. Despite differences in the Reynolds number, the velocity profiles now
agree very well, throughout the outer layer.
This observation would suggest problems with correct estimation of the friction velocity, which however seems unlikely both in DNS,
in which we independently evaluate friction velocity by computing the wall derivative of the velocity profile
and from momentum balance, and in experiments, as measurements of the pressure drop are regarded
to have low uncertainty.
Hence, reasons for this discrepancy are not known, and additional experiments as those currently
carried out in the large CICLoPE facility would be especially useful and welcome. 
Unfortunately, velocity profiles along the full radial span are not available at the moment for that facility.

\begin{figure}
 \centerline{
(a)~\includegraphics[width=7.0cm]{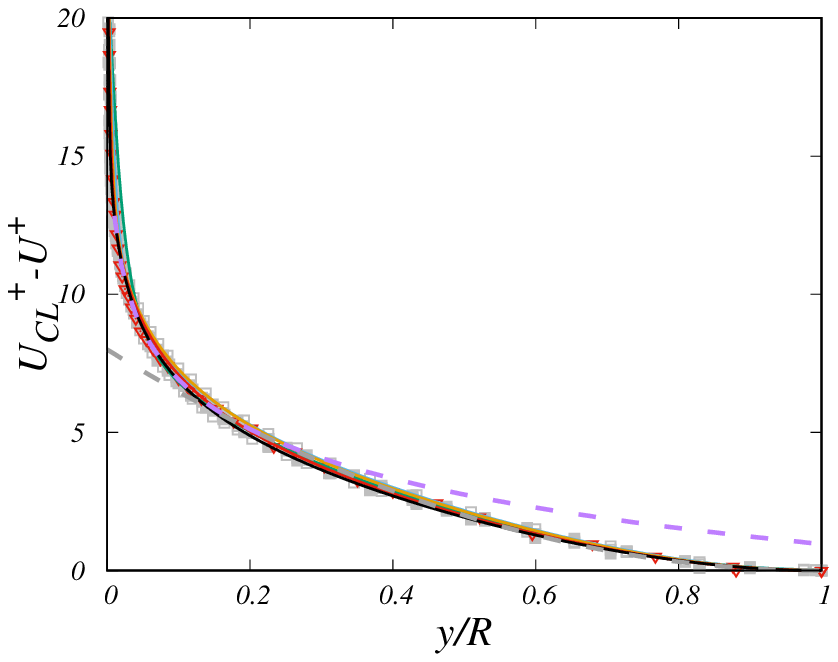}~(b)~\includegraphics[width=7.0cm]{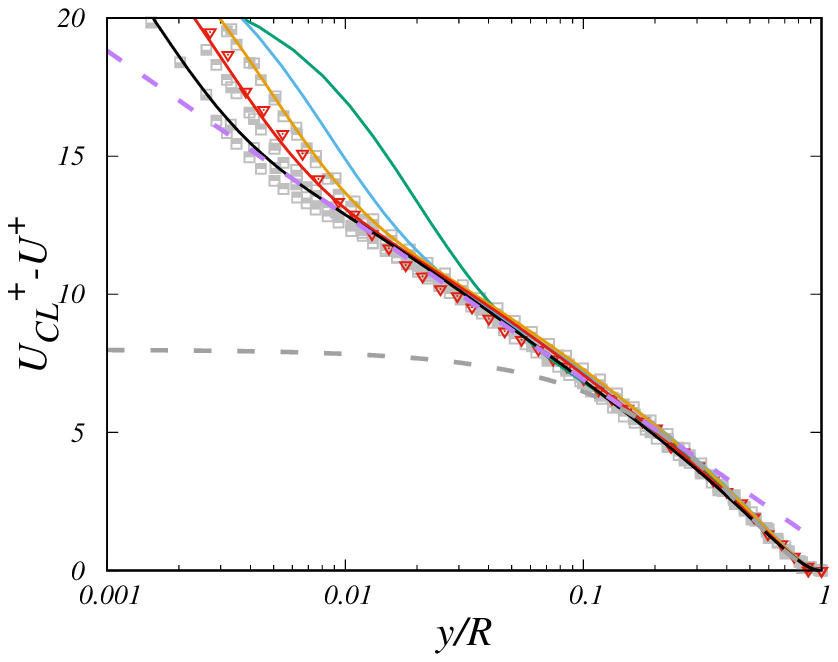}
 }
\caption{Defect velocity profiles for DNS and experiments, in linear (a) and semi-logarithmic (b) scale. 
The dashed grey line marks a parabolic fit of the DNS data ($U^+_{_{CL}}-U^+ = 8.0 (1-y/R)^2$),
and the dashed purple line the outer-layer logarithmic fit $U^+_{_{CL}}-U^+ = 0.961 - 1/0.387 \log (y/R)$.}
\label{fig:defect}
\end{figure}

The structure of the core region is examined in detail in figure~\ref{fig:defect}, where the mean velocity profiles are shown in defect form.
Although full outer-layer similarity is not reached at the conditions of our DNS study
(also see the inset of figure~\ref{fig:uplus}(a)), scatter across the Reynolds number range and 
with respect to Superpipe profiles for $y/R \ge 0.2$ is no larger than $5\%$. 
As suggested by \citet{pirozzoli_14a}, the core velocity profiles can be closely approximated 
with a simple quadratic function, reflecting near constancy of the eddy viscosity.
In particular, we find that the formula
\begin{equation}
U_{_{CL}}^+ - U^+ = C_{_{O}} \left( 1 - y/R \right)^2, \label{eq:parabola}
\end{equation}
fits the DNS data with $C_{_{O}} = 8.0$ well, and it smoothly connects at $y/R \approx 0.2$ with the 
logarithmic profile expressed in outer form,
\begin{equation}
U_{_{CL}}^+ - U^+ = - \frac 1k \, \log (y/R) + B , \label{eq:loglaw_out}
\end{equation}
where again $k=0.387$, and data fitting yields $B=0.961$. While of course better descriptions of the 
core velocity profiles are possible based on more elaborate functional relationships~\citep{luchini_17},
the composite profile matching equations~\eqref{eq:parabola} and \eqref{eq:loglaw_out} yields 
a reasonable representation of the whole outer-layer mean velocity profile within the 
scatter of available data.

\begin{figure}
 \centerline{
(a)~\includegraphics[width=7.0cm]{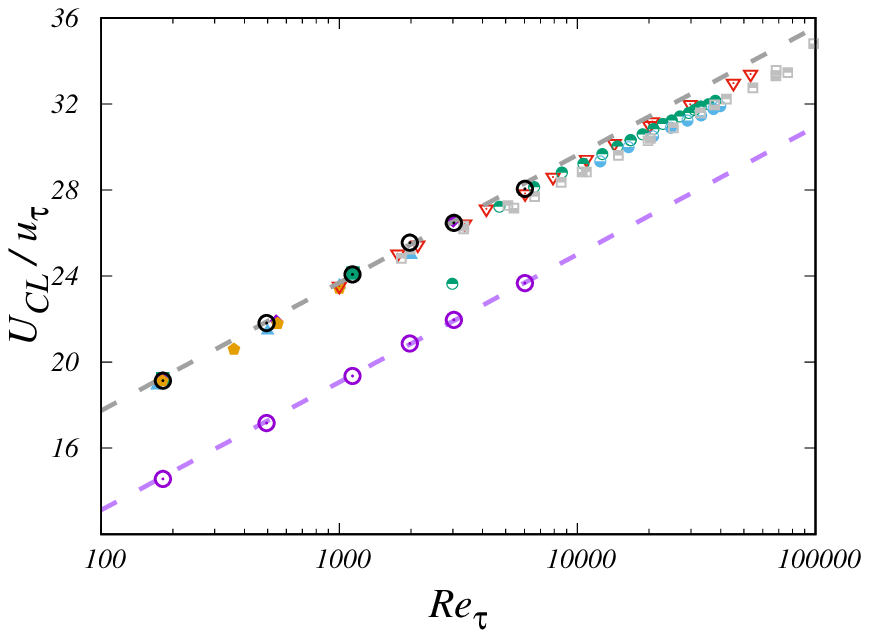}~(b)~\includegraphics[width=7.0cm]{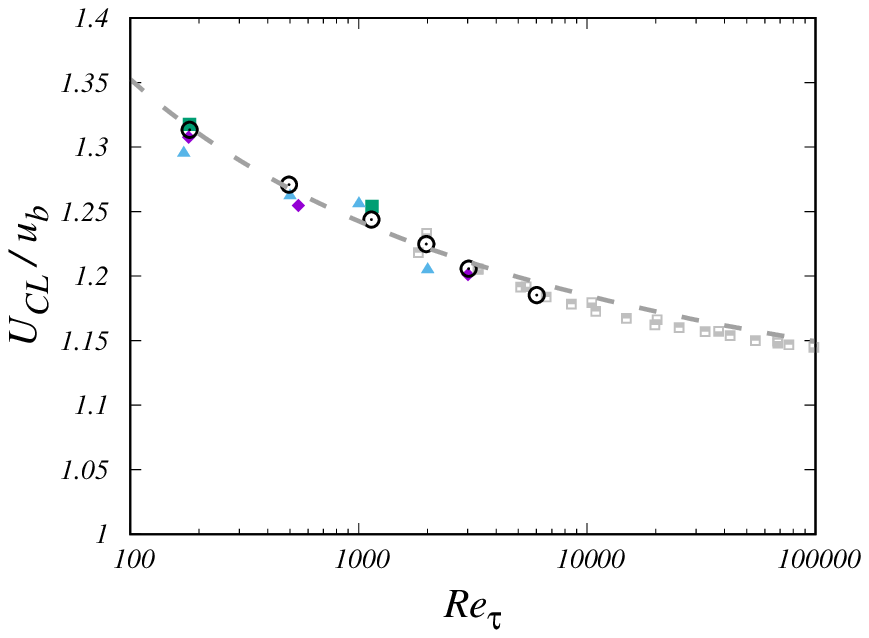}}
\caption{Mean pipe centerline velocity ($U_{_{CL}}$) expressed in inner (a) and in outer (b) units.
The dashed grey line corresponds to a fit of the DNS data. 
DNS data are shown as circle symbols, and the corresponding logarithmic fits are shown as thick dashed lines. 
Purple lines and symbols are used for the bulk velocity, $u_b$. 
For the nomenclature of other symbols, refer to table~\ref{tab:refs}.}
\label{fig:ucl}
\end{figure}


Finer evaluation of similarities and differences between DNS and experiments is provided in figure~\ref{fig:ucl},
where we show the mean centerline velocity, $U_{_{CL}}$, normalized by the friction velocity (left panel),
and by the bulk velocity (right panel), as a function of the friction Reynolds number. 
Consistently with theoretical expectations~\citep[e.g.][]{monkewitz_21}, data suggest logarithmic increase 
with $\Rey_{\tau}$ according to 
\begin{equation}
U_{_{CL}}^+ = \frac 1{k_{_{CL}}} \log \Rey_{\tau} + B_{_{CL}},
\end{equation}
where we find $k_{_{CL}}=k=0.387$ as for the friction law, and $B_{_{CL}}=5.85$. For convenience, the trend of $u_b/u_{\tau}$ is also
presented, having in fact the same logarithmic growth with $\Rey_{\tau}$.
With some previously noted differences, all pipe flow DNSs seem to exhibit a consistent trend in the accessible range.
While the trend is very similar at low Reynolds number, experimental data yield consistently lower values of $U_{_{CL}}^+$,
especially those from the Superpipe. At Reynolds numbers higher than about $\Rey_{\tau}=10^4$, 
experiments seem to suggest milder growth rate, although significant differences emerge between the Superpipe
and the Hi-Reff datasets. Hence, whether this is the result of a change of behaviour at high
Reynolds number, or some form of shortcoming of experiments is difficult to say.
As a result of the observed identity (or very close vicinity) of the von K\'arm\'an constant for 
the centerline and for the bulk velocity, 
figure~\ref{fig:ucl}(b) highlights that their ratio approaches unity at large $\Rey$,
supporting the inference that pipe flow asymptotes to plug flow in the infinite-Reynolds-number limit~\citep{pullin_13}.
Regarding that study, it is worthwile noticing that one of the assumptions made in the
analysis is that the wall-normal location of the onset of the logarithmic region is either finite,
or increases no faster than $\Rey_{\tau}$. Interpreting the near-wall minimum of the diagnostic function
in figure~\ref{fig:diag} as the root of the (near) logarithmic layer, our data well support that assumption.
Whereas the curvature of the core velocity profile is not changing substantially 
when expressed in wall units (see figure~\ref{fig:defect}), it would become vanishingly small when expressed in outer units.
However, as figure~\ref{fig:ucl}(b) suggests, this trend is extremely slow.
Interestingly, again despite some scatter, DNS and experiments here seem to indicate a common trend
with overall monotonic decrease, perhaps with a 'bump' in the range of Reynolds numbers in the 
low thousands. The DNS data points at the highest Reynolds numbers (DNS-D,E,F) now appear to be 
in good agreement with Superpipe experiments, which is in line with the previously noted
agreement of the outer-scaled mean velocity profiles.

\begin{figure}
 \centerline{
 (a)~\includegraphics[width=7.0cm]{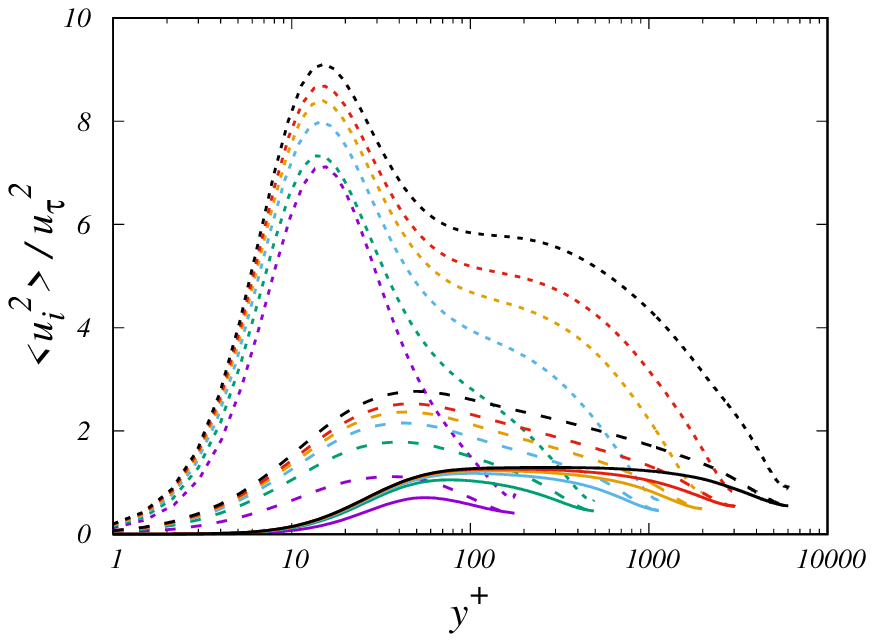}~(b)~\includegraphics[width=7.0cm]{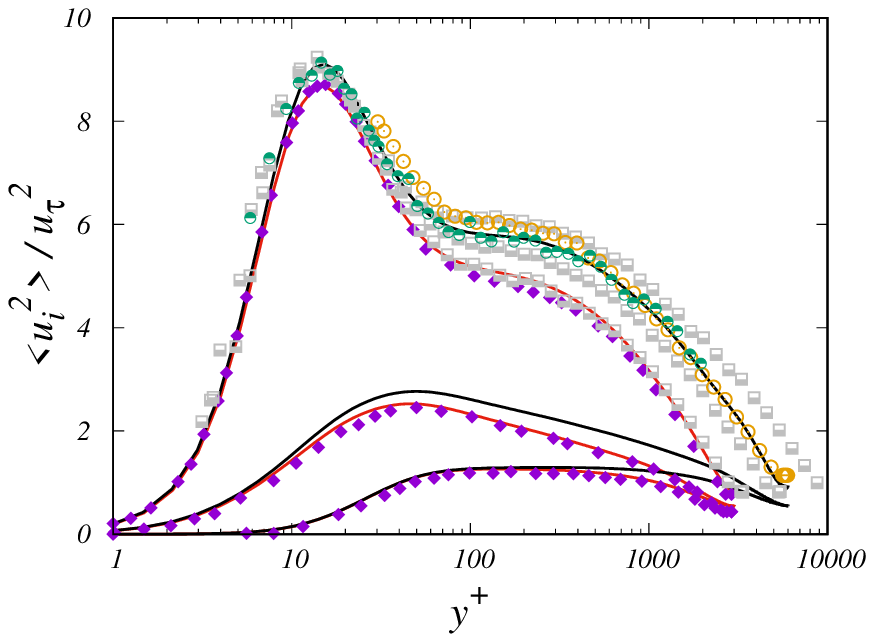}} 
 \caption{Distribution of velocity variances (a) and comparison of cases DNS-E, DNS-F with reference DNS and experiments (b).
In panel (a), the short dashed lines denote the axial velocity variance ($< u_z^2 >$),
the solid lines denote the radial velocity variance ($< u_r^2 >$),
and the long dashed lines denote the azimuthal velocity variance ($< u_{\theta}^2 >$). 
For color codes in DNS data, see table~\ref{tab:runs}, and for nomenclature of symbols, see table~\ref{tab:refs}.}
\label{fig:urms}
\end{figure}

The distributions of the velocity variances along the coordinate directions
are shown in figure~\ref{fig:urms}, in inner scaling. As now well established~\citep{marusic_19}, 
the longitudinal ($u_z$) and spanwise ($u_{\theta}$) velocity fluctuations show slow increase with the Reynolds number,
with commonly accepted logarithmic growth as after Townsend's attached eddy model~\citep{townsend_76}.
On the other hand, the wall-normal velocity fluctuations seem to level off to a maximum value of about $1.30$.
It is remarkable that the general growth of the longitudinal and spanwise fluctuations
is more evident in the outer layer, and in fact it has long been argued about the possible
occurrence of a secondary peak of $<u^2_z>$, besides the primary buffer-layer peak.
Experiments carried out in the Superpipe~\citep{hultmark_12} and CICLoPE~\citep{willert_17} facilities
indeed support the occurrence of such peak at $\Rey_{\tau} \gtrsim 10^4$.
Whereas DNS data are not at sufficiently high $\Rey_{\tau}$ 
to show this secondary peak, it appears that in DNS-F the axial velocity variance has attained
a nearly horizontal inflectional point at $y^+ \approx 140$. 
Comparison with the $\Rey_{\tau} \approx 3000$ DNS of \citet{ahn_15}
shows overall good agreement of all turbulence intensities.
Comparison with Superpipe data at $\Rey_{\tau} = 3000$ is also very good, with exception of the
near-wall peak which is likely to be over-estimated in experiments. 
DNS-F data seem to be well bracketed by Superpipe and CICLoPE measurements at nearby Reynolds numbers,
and also compare very well with experimental data for plane channel flow~\citep{flack_13}. 

\begin{figure}
 \centerline{
 (a)~\includegraphics[width=7.0cm]{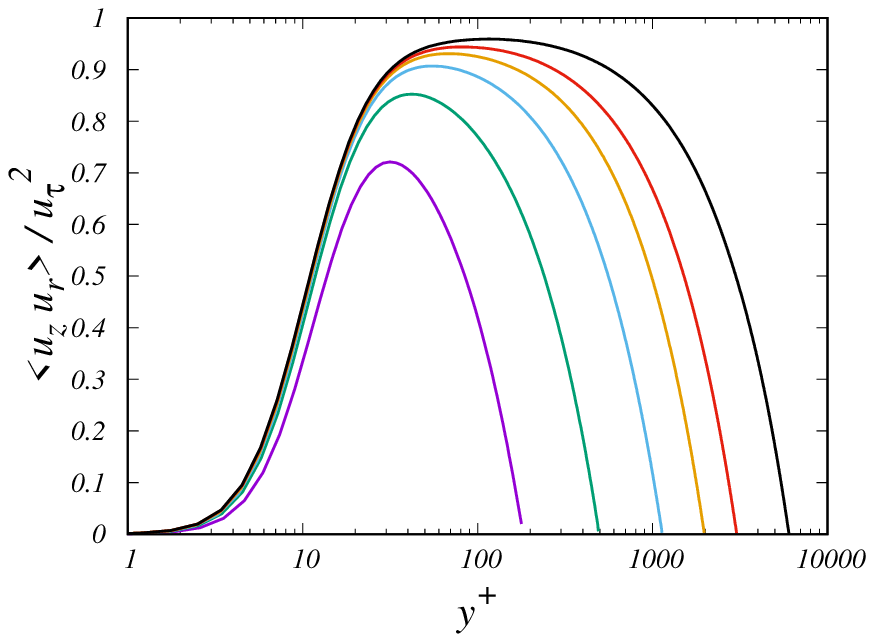}~(b)~\includegraphics[width=7.0cm]{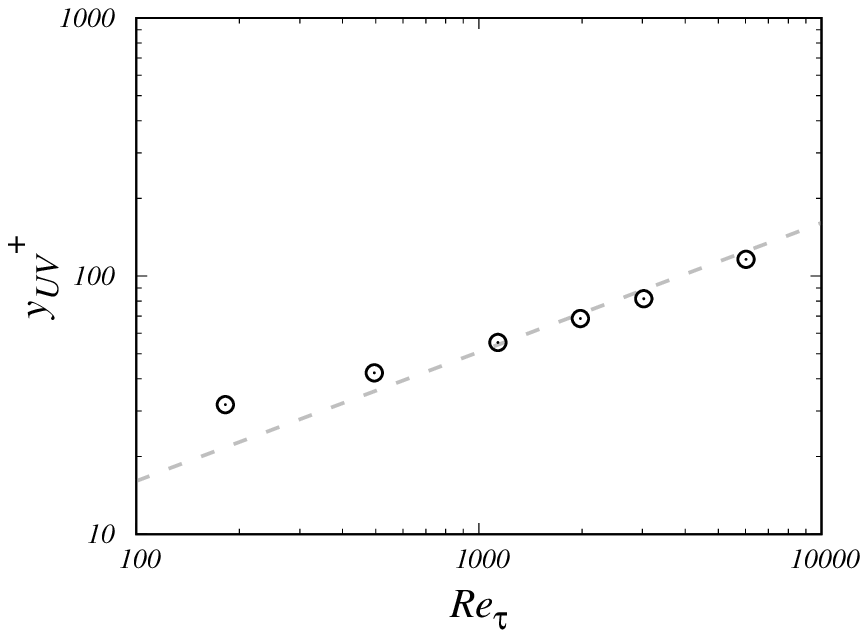}} 
 \caption{Distributions of turbulent shear stress (a) and its peak position at various $\Rey_{\tau}$ (b).
The dashed line in panel (b) denotes the theoretical estimate~\eqref{eq:uvpeak}.
For color codes in DNS data, see table~\ref{tab:runs}.}
\label{fig:uv}
\end{figure}

Distributions of the turbulent shear stress are shown in figure~\ref{fig:uv}. As is well established~\citep[e.g.][]{lee_15}, 
the shear stress profiles tend to become flatter at higher $\Rey_{\tau}$, the peak value rises towards unity, and
its position moves farther from the wall, in inner units. In particular, exploiting mean momentum balance and
assuming the presence of a logarithmic layer in the mean axial velocity, the following prediction follows 
for the position of the turbulent shear stress peak~\citep{afzal_82}
\begin{equation}
y_{_{UV}}^+ \approx \sqrt{\frac{\Rey_{\tau}}k}, \label{eq:uvpeak}
\end{equation}
which is intermediate between inner and outer scaling. This observation has led some authors
to argue about the relevance of a 'mesolayer'~\citep{long_81,wei_05}.
The asymptotic relationship \eqref{eq:uvpeak} is satisfied with good accuracy starting at $\Rey_{\tau} \approx 10^3$,
reflecting the onset of a near logarithmic layer.

\begin{figure}
 \centerline{
 (a)~\includegraphics[width=7.0cm]{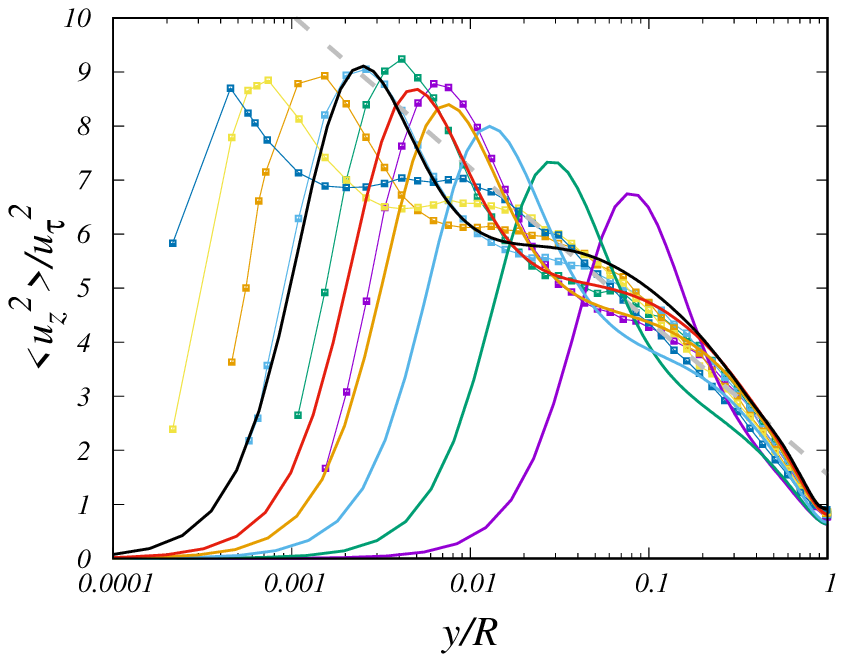}~(b)~\includegraphics[width=7.0cm]{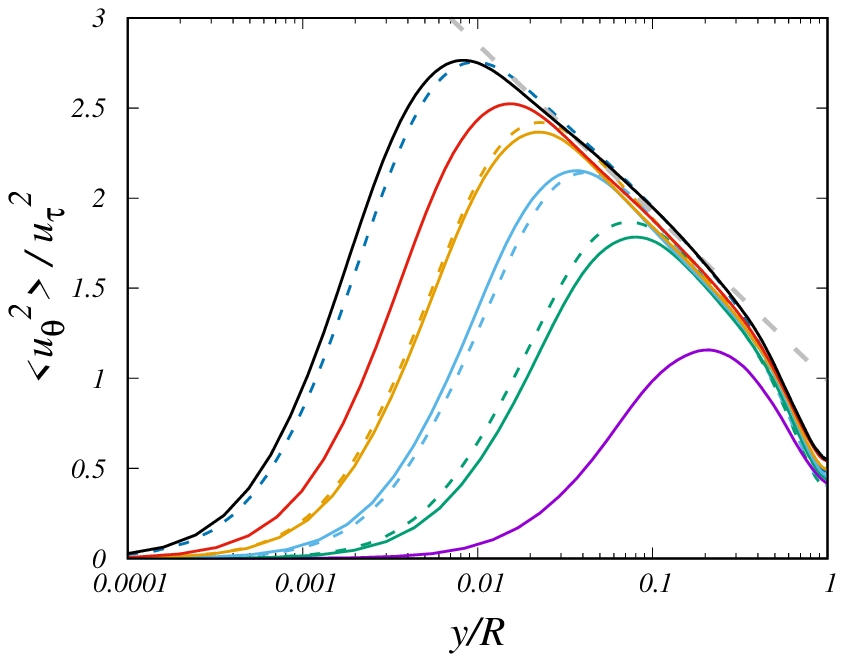}} 
 \caption{Axial (a) and azimuthal (b) turbulent stresses as a function of outer-scaled wall distance.
In panel (a), symbols denote Superpipe data~\citep{hultmark_12} at 
$\Rey_{\tau} = 1985, 3334, 5411, 10480, 20250, 37690$, and the 
dashed grey line the corresponding fit, $\left< u_z^2 \right> = 1.61-1.25 \log (y/R)$.
In panel (b), the dashed colored lines denote DNS data of channel flow~\citep{lee_15}
at $\Rey_{\tau} = 550, 1000, 2000, 5200$, and the dashed grey line the fit of the DNS data,
$\left< u_{\theta}^2 \right> = 1.0 - 0.40 \log (y/R)$.
For color codes in DNS data, see table~\ref{tab:runs}.}
\label{fig:urms2_outer}
\end{figure}

The behaviour of the Reynolds stresses when expressed as a function of the outer-scaled wall distance,
which is shown in figure~\ref{fig:urms2_outer} is also of great theoretical interest.
In fact, according to the attached-eddy model~\citep{townsend_76, marusic_19}, the wall-parallel velocity variances 
are expected to decline logarithmically with the wall distance in the outer layer, hence
\begin{equation}
\left< u_z^2 \right>        = B_1 - A_1 \log (y/R), \quad
\left< u_{\theta}^2 \right> = B_3 - A_3 \log (y/R). 
\end{equation}
where $A_i$, $B_i$ are universal constants. 
Regarding the axial stress, \citet{marusic_13} argued that Superpipe data 
at the highest available Reynolds number are best fit with $A_1=1.23$, $B_1=1.56$,
with a sensible logarithmic layer only emerging at $\Rey_{\tau} > 10^4$,
in the range of wall distances $3 \Rey_{\tau}^{1/2} \le y^+ \le 0.15 \Rey_{\tau}$.
DNS data only show the formation of a near logarithmic layer farther away from the wall,
which is not where it is expected from theoretical arguments. Hence, little 
can be said in this respect.
The azimuthal velocity variance, shown in figure~\ref{fig:urms2_outer}(b), has a 
more benign behaviour, and it features clear logarithmic layers even at modest $\Rey_{\tau}$.
Fitting the DNS data yields $A_3 = 0.40$, $B_3 = 1.0$, which is very close to what found 
in channels~\citep{bernardini_14, lee_15}. Measurements of pipe flow carried out in the
CICLoPE facility~\citep{orlu_17} yielded $A_3 = 0.63$, $B_3=1.21$, hence
much larger values than in DNS. Possible overestimation of the wall-normal and azimuthal 
Reynolds stresses was in fact acknowledged by the authors of that paper.

\begin{figure}
 \centerline{
(a)~\includegraphics[width=7.0cm]{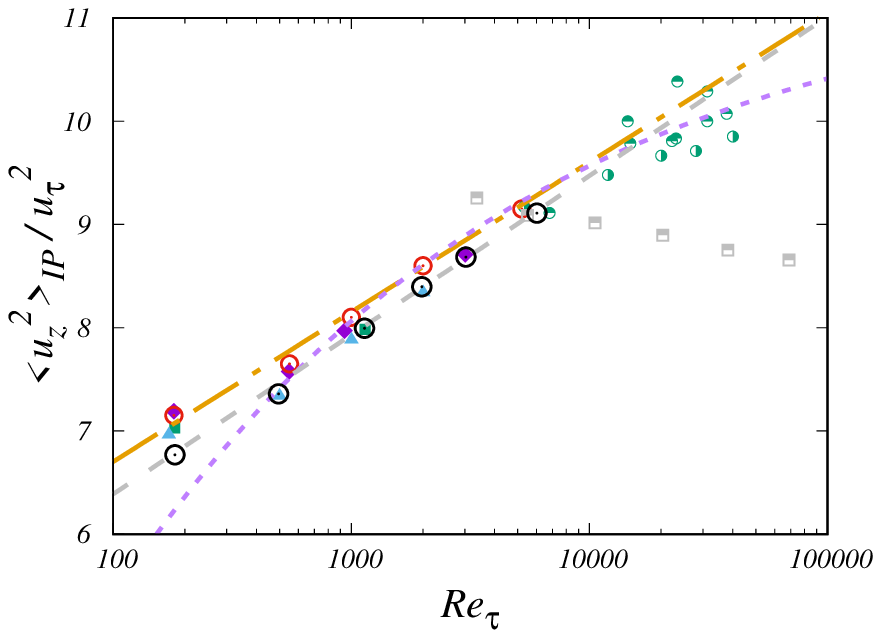}~(b)~\includegraphics[width=7.0cm]{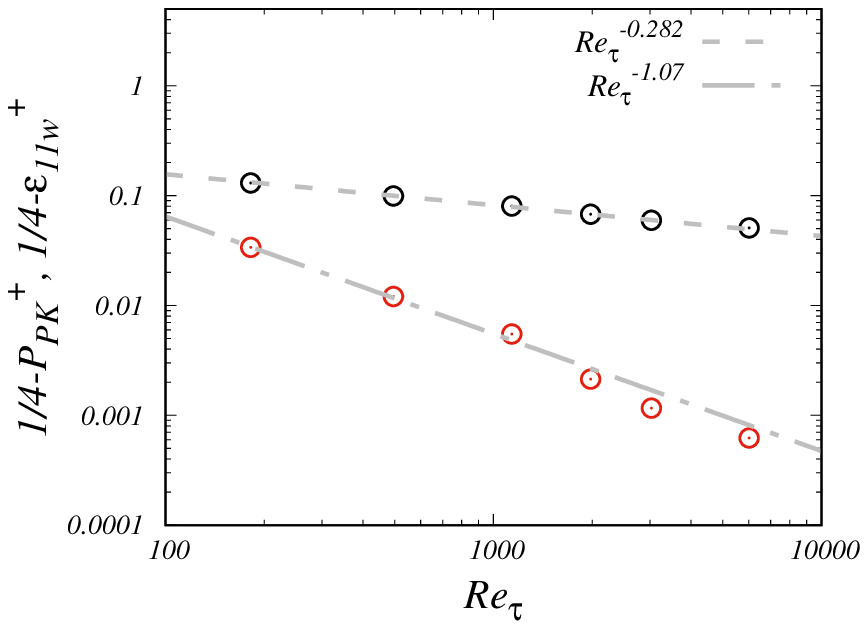}}
\caption{Magnitude of inner peak of axial velocity variance (a) and peak turbulence production ($P_{_{PK}}$, red), and wall dissipation of axial velocity variance ($\epsilon_{_{11w}}$, black) (b). For color codes in DNS data, see table~\ref{tab:runs}, and for nomenclature of symbols, see table~\ref{tab:refs}.
In panel (a) the dashed grey line marks the DNS data fit, $<u^2_z>^+_{_{IP}} = 0.67 \log \Rey_{\tau} + 3.3$,
the dashed purple line denotes the defect power law of \citet{chen_21}, and the dash-dotted line the logarithmic 
law of \citet{marusic_17}, $<u^2_z>^+_{_{IP}} = 0.63 \log \Rey_{\tau} + 3.8$. 
In panel (b), the dot-dashed and dotted lines denote fits of $P_{_{PK}}$ and $\epsilon_{_{11w}}$ 
in their tendency to the respective assumed asymptotic values.}
\label{fig:innerpeak}
\end{figure}

Quantitative insight into Reynolds number effects is provided by inspection of the amplitude of the inner peak of the axial velocity variance,
which we show in figure~\ref{fig:innerpeak}. 
The general theoretical expectation is that the peak grows logarithmically with $\Rey_{\tau}$ owing to the 
increasing influence of distant, inactive eddies~\citep{marusic_19}. However, some recent experimental results~\citep{willert_17},
and theoretical arguments~\citep{chen_21} suggest that such growth should eventually saturate.
Although difference between slow logarithmic growth and attainment of an asymptotic value is quite subtle in practice,
the theoretical interest is high as in the latter case universality of wall scaling would be eventually restored.
Within the investigated range of Reynolds numbers, our DNS data in fact support continuing logarithmic increase.
Comparison with channel data~\citep{lee_15} shows some difference, which might result
from stronger geometrical confinement of distant eddies in the pipe geometry.
However, differences tend to becomes smaller at higher $\Rey_{\tau}$. 
In quantitative terms, we find the slope of logarithmic increase to be about $0.67$,
a bit steeper than found in channel flow DNS~\citep[about $0.64$]{lee_15}, and than
suggested from a collection of DNS and experiments~\citep[about $0.63$]{marusic_17}.
Experimental data for pipe flow are quite scattered, as Superpipe experiments
yield an unrealistically decreasing trend~\citep{hultmark_12}, PIV measurements taken in the CIPLoPE facility~\citep{willert_17}
suggest saturation of the growth, whereas hot-wire measurements in the same facility support
continued logarithmic growth~\citep{fiorini_17}. 
The theoretical predictions of \citet{chen_21} (see the dashed purple line of figure~\ref{fig:innerpeak}a) 
seem to conform well with channel flow DNS data and with the experiments of \citet{willert_17}. 

While our DNS data cannot be used to directly evaluate the theoretical predictions owing to limited achievable
Reynolds number, they can be used to better scrutinize the foundations of the theoretical arguments. 
The main argument made by \citet{chen_21}, although not thoroughly justified in our opinion, was 
that since turbulence kinetic energy production is bounded, the wall dissipation must also 
stay bounded. Hence, let $P = - \left< u_z u_r \right> \diff U / \diff r$ be the 
turbulence kinetic energy production rate, and $\epsilon_{11} = \nu \left< \left| \nabla u_z \right|^2 \right>$
be the dissipation rate of the axial velocity variance, those authors first argue that the wall limiting 
value of $\epsilon_{11}$ should scale as
\begin{equation}
{\epsilon_{11}}_w^+ = 1/4 - \beta / \Rey_{\tau}^{1/4}, \label{eq:eps1w}
\end{equation}
with $\beta$ a suitable constant. In figure~\ref{fig:innerpeak} we explore deviations of 
$\epsilon_w$ and of the peak turbulence kinetic energy production, say $P_{_{PK}}$, from their
asymptotic value, namely $1/4$. According to analytical constraints~\citep{pope_00}, we find that
production tends to its asymptotic value quite rapidly, as about $1/\Rey_{\tau}$.
Consistent with equation~\eqref{eq:eps1w}, the wall dissipation also tends to $1/4$, more or less
at the predicted rate, thus empirically validating the first assumption.
The next argument advocated by \citet{chen_21} is that wall balance between viscous diffusion 
and dissipation and Taylor series expansion of the axial velocity variance near the wall yields
\begin{equation}
\left< u_z^2 \right>^+ \sim \epsilon_{_{11w}}^+ {y^+}^2, \label{eq:taylor}
\end{equation}
whence, from assumed invariance of the peak location of $\left< u_z^2 \right>$ (say, $y_{_{IP}}^+$),
saturation of growth of the peak velocity variance would follow. 
Table~\ref{tab:UQ} suggests that this second assumption is in fact violated, as the position of 
the peak slightly increases with $\Rey_{\tau}$, with non-negligible effect on the peak variance
as it appears in squared form in equation~\eqref{eq:taylor}.
As a consequence, logarithmic growth of the peak velocity variance still holds, 
at least in the range of Reynolds numbers currently accessible to DNS.

\begin{figure}
 \centerline{
 (a)~\includegraphics[width=7.0cm]{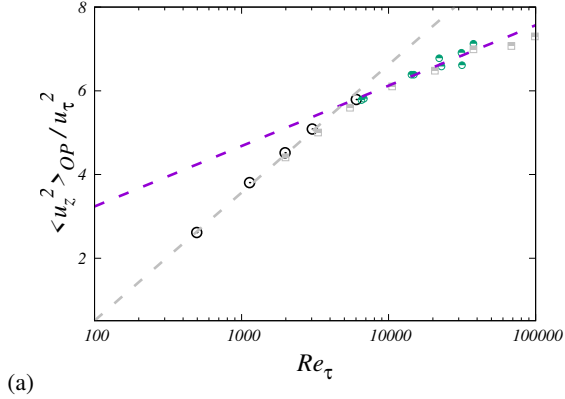}}
\caption{Magnitude of outer peak of axial velocity variance as a function of $\Rey_{\tau}$. 
Lines and symbols as in tables~\ref{tab:runs} and \ref{tab:refs}.
The dashed grey line marks the DNS data fit, $\left< u_z^2 \right>^+_{OP} = 1.33 \log \Rey_{\tau} - 5.61$,
and the purple line denotes the logarithmic fit given by \citet{pullin_13}. 
}
\label{fig:outerpeak}
\end{figure}

Although no distinct outer peak of the axial velocity variance is found at the Reynolds numbers under scrutiny, 
it is nevertheless instructive to explore the scaling of the velocity fluctuations in the range of wall
distances where the peak is expected to occur. For that purpose, we
consider the outer position where the second logarithmic derivative of the velocity variance vanishes,
which in the present DNS ranges from $y^+ \approx 115$ for DNS-A, to $y^+ \approx 140$ for DNS-F.
Weak dependence of the inner-scaled outer peak position on $\Rey_{\tau}$,
although at much higher Reynolds number, was also noticed by \citet{hultmark_12}.
The resulting distribution is shown in figure~\ref{fig:outerpeak}. All DNS data fall nicely on a logarithmic 
fit, and they seem to connect smoothly to the experimental results, whose scatter and
uncertainty is expected to be much less than for the inner peak. Experiments indicate a change of 
behaviour to a shallower logarithmic dependence with slope of about $0.63$~\citep{fiorini_17, pullin_13},
which would be very close to the growth rate of the inner peak (see figure~\ref{fig:innerpeak}).
The figure suggests that verification of this effect would require $\Rey_{\tau}$ of about $10^4$.

\begin{figure}
 \centerline{
 (a)~\includegraphics[width=7.0cm]{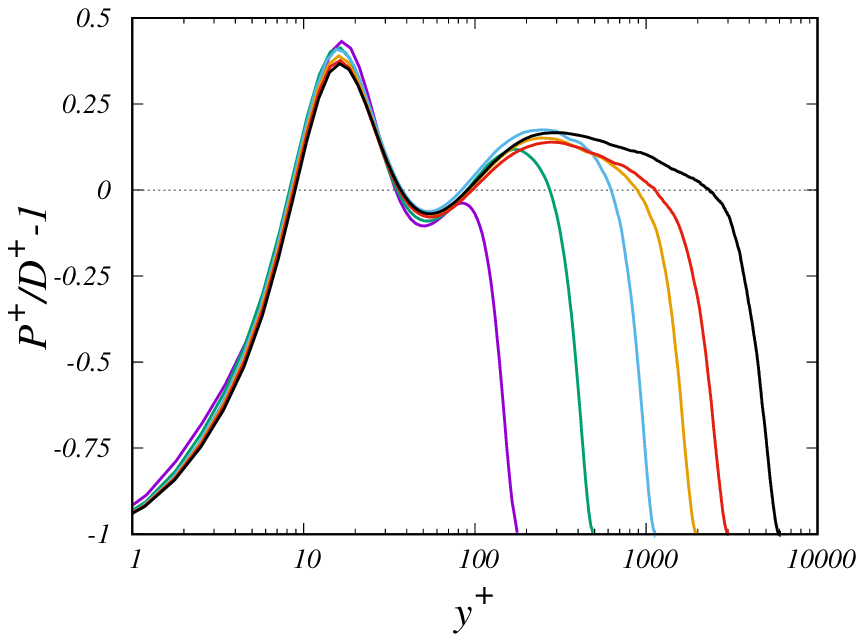}~(b)~\includegraphics[width=7.0cm]{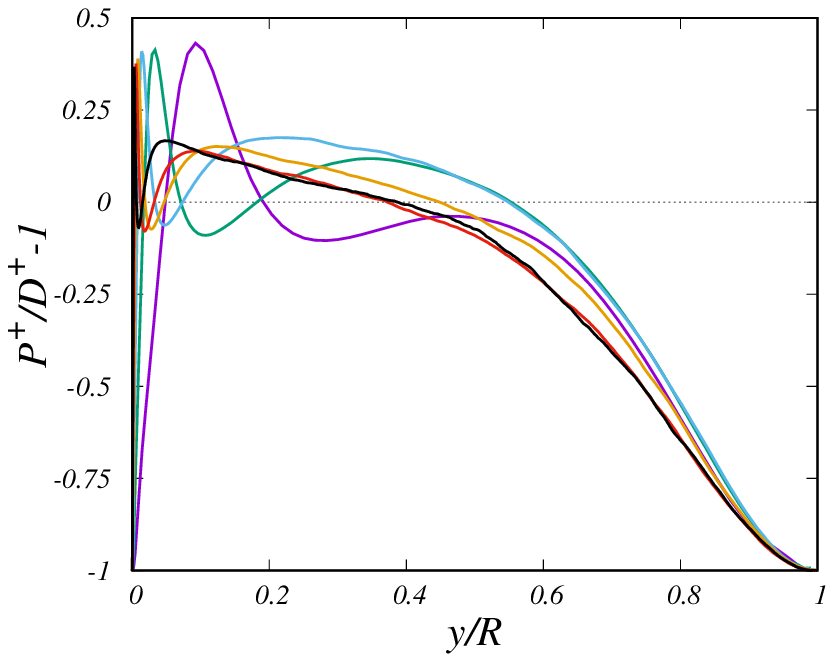}}
\caption{Excess of turbulence kinetic energy production over dissipation as a function of inner-scaled (a)
and outer-scaled (b) wall distance. Lines as in table~\ref{tab:runs}.}
\label{fig:p_by_eps}
\end{figure}

As pointed out by \citet{hultmark_12}, the formation and growth of an outer peak of 
the axial velocity variance has important theoretical and practical implications. 
From the modeling standpoint, no current RANS model is capable of predicting 
non-monotonic behaviour of Reynolds stresses outside the buffer layer. From the fundamental physics standpoint,
the presence of an outer peak is suggestive of violation of equilibrium 
between turbulence production and dissipation, thus invalidating 
one of the possible arguments advocated for
the formation of a logarithmic mean velocity profile~\citep{townsend_76}. 
DNS allows to substantiate this scenario, and for that purpose in figure~\ref{fig:p_by_eps}, we show the 
relative excess of turbulent kinetic energy production ($P$) over its total
dissipation rate, here defined as 
$D = \nu \left< u_i \nabla^2 u_i \right>$, which lumps together dissipation rate and viscous diffusion.
Data confirm the presence of a near-universal region confined to the buffer layer 
(say, $8 \lesssim y^+ \lesssim 35$), in which production exceeds dissipation by up to $40\%$.
Data also show the onset, starting from DNS-B, of another region farther from the wall with positive unbalance,
whose inner limit is constant in inner units, at $y^+ = 100$,
and whose outer limit tends to become constant at high $\Rey_{\tau}$ in outer units, at $y/R \approx 0.4$.
The peak unbalance at high Reynolds number is about $17\%$, and its position seems 
to scale more in inner than in outer units. 
Turbulence kinetic energy production excess in the presence of a (near) logarithmic mean velocity profile
can be interpreted by recalling that only part of the kinetic energy which is being generated 
is converted to active motions which carry turbulent shear stress, 
and the rest is used to feed inactive motions.
This finding clearly indicates that at high enough Reynolds number the outer wall layer becomes 
a dynamically active part of the flow, having the potential to transfer energy both to the 
core flow, and towards the wall, in the form of imprinting on the near-wall layer~\citep{marusic_19}. 



\section{Concluding comments}

Although DNS of wall turbulence is still confined to a moderate range of Reynolds numbers, 
it is beginning to approach a state in which some typical phenomena of the 
asymptotically high-$\Rey$ emerge. Given its ability to resolve the near-wall layer,
DNS lends itself to testing theories of wall turbulence and to in-depth scrutiny of experimental data.
In this work, DNS of flow in a smooth pipe has been carried out up to $\Rey_{\tau} \approx 6000$,
which, although still far from what achievable in experimental tests, allows to uncover a number of 
interesting issues, in our opinion.
First, we have noted that DNS data fall systematically short of
the classical Prandtl friction law, by as much as $2\%$. This evidence is not consistent with 
data from the Superpipe facility, although other recent data from the CICLoPE and Hi-Reff
facilities seem to yield similar trends. DNS data fitting suggests that a
logarithmic law as \eqref{eq:prandtl} still holds, however with a von K\'arm\'an constant $k \approx 0.387$,
which matches extremely well the value quoted by \citet{furuichi_18}, and which 
would reconcile pipe flow with plane channel and boundary layer flows,
thus corroborating claims made by \citet{marusic_13}.
A logarithmic profile with $k \approx 0.387$ well fit the mean axial velocity distributions
for $30 \le y^+ \le 0.15 \Rey_{\tau}$,
although linear deviations are clearly visible, as argued by \citet{afzal_73, luchini_17},
which is taken into account yield excellent representation of the
velocity profiles up to $y/R \approx 0.5$. 
It is remarkable that the same value of the von K\'arm\'an constant also well fits the mean centerline velocity
distribution (see figure~\ref{fig:ucl}), which is found to grow logarithmically throughout the
range of $\Rey_{\tau}$ under investigation. This finding is quite reasonable as it corroborates that
the eventual state of turbulent flow in a pipe should be plug flow, as argued by \citet{pullin_13},
hence $U_{_{CL}} \to u_b$ as $\Rey_{\tau} \to \infty$. This would however seemingly contrast recent
measurements made in the CICLoPE facility~\citep{nagib_17}, which rather suggest 
a different von K\'arm\'an constant for the bulk and the centerline velocity. 
Experimental data at $\Rey_{\tau} \gtrsim 10^4$ in fact suggest 
deviations of $U_{_{CL}}^+$ from the logarithmic trend found DNS, 
however this effect requires further confirmation, as data are quite scattered.
The core velocity profile is found to be to a good approximation parabolic, 
with curvature which is nearly constant in wall units, and decreasing in outer units.

Regarding the velocity fluctuations, we find evidence for continuing logarithmic increase of the inner-peak
magnitude with $\Rey_{\tau}$. Some experiments and theoretical arguments would indicate that
beyond $\Rey_{\tau} \approx 10^4$ a change of behaviour might occur, which however is very difficult to quantify. 
DNS is probably of little use in this respect, as in order to clearly discern among the various 
trends, $\Rey_{\tau}$ in excess of $10^5$ are likely to be needed. 
As predicted by the attached-eddy hypothesis, the wall-parallel velocity variances in the outer layer
tend to form logarithmic layers, which are especially evident in the azimuthal velocity.
Although we do not find direct evidence for
the existence of an outer peak of the axial velocity variance, our results highlight the 
occurrence of an outer site with substantial turbulence production excess over dissipation,
thus contradicting the classical equilibrium hypothesis and likely to yield 
a distinct peak at $\Rey_{\tau} \approx 10^4$, as Superpipe and CICLoPE experiments suggest.
Investigating these and other violations of universality of wall turbulence to extrapolate asymptotic behaviours
is a formidable challenge for theoreticians in years to come.



\backsection[Acknowledgments]{
We acknowledge that the results reported in this paper have been achieved using the PRACE Research Infrastructure resource MARCONI based at CINECA, Casalecchio di Reno, Italy, under project PRACE n. 2019204979. Discussions with A.J. Smits are gratefully acknowledged.
We would like to thank P. Luchini and M. Quadrio for providing the code used for the data uncertainty analysis.}

\backsection[Funding]{This research received no specific grant from any funding agency, commercial or not-for-profit sectors.}

\backsection[Declaration of interests]{The authors report no conflict of interest.}

\backsection[Data availability statement]{The data that support the findings of this study are openly available at 
the web page http://newton.dma.uniroma1.it/database/}



%
\bibliographystyle{jfm}
\bibliography{references}
\end{document}